\documentclass[letterpaper,twocolumn,pra,aps,showpacs,superscriptaddress,floatfix]{revtex4-1}
\usepackage[latin1]{inputenc}
\usepackage{bm}
\usepackage[usenames]{color}
\usepackage{multirow}
\usepackage{amssymb}
\usepackage{amsbsy}
\usepackage{amsmath}
\usepackage{stmaryrd}
\usepackage{graphicx}
\usepackage{epsfig}
\usepackage{placeins}
\usepackage{ulem}
\makeatletter
\begin{document}

\title{Scaling of diffusion constants in the spin-$1/2$ XX ladder}

\author{R. Steinigeweg}
\email{r.steinigeweg@tu-bs.de}
\affiliation{Institute for Theoretical Physics,
Technical University Braunschweig, D-38106 Braunschweig, Germany}

\author{F. Heidrich-Meisner}
\affiliation{Department of Physics and Arnold Sommerfeld Center for Theoretical Physics,
Ludwig-Maximilians-Universit\"at M\"unchen, D-80333 M\"unchen, Germany}

\author{J. Gemmer}
\affiliation{Department of Physics,
University of Osnabr\"uck, D-49069 Osnabr\"uck, Germany}

\author{K. Michielsen}
\affiliation{Institute for Advanced Simulation, J\"ulich Supercomputing Centre,
Forschungszentrum J\"ulich, D-52425 J\"ulich, Germany
and RWTH Aachen University, D-52056 Aachen, Germany}

\author{H. De Raedt}
\affiliation{Department of Applied Physics, Zernike Institute for Advanced Materials,
University of Groningen, NL-9747AG Groningen, The Netherlands}

\date{\today}

\begin{abstract}
We study the dynamics of spin currents in the XX spin-$1/2$ ladder
at finite temperature. Within  linear response theory, we
numerically calculate autocorrelation functions for quantum systems
larger than what is accessible with exact diagonalization using the
concept of dynamical quantum typicality. While the spin Drude weight
vanishes exponentially fast with increasing system size, we show
that this model realizes standard diffusive dynamics. Moreover, we
unveil the existence of three qualitatively different dependencies
of the spin diffusion coefficient on the rung-coupling strength,
resulting from a crossover from exponential to Gaussian dissipation
as the rung coupling increases, in agreement with analytical
predictions. We further discuss the implications of our results for
experiments with cold atomic gases.
\end{abstract}


\maketitle

\section{Introduction}
\label{sec:intro}

The theoretical understanding of transport
properties of interacting quantum many-body systems is paramount in
characterizing states of matter. Strongly interacting
one-dimensional (1D) systems may exhibit either diffusive or
ballistic transport properties at finite temperatures
\cite{zotos1997, hm07, znidaric2011, steinigeweg2011-2}, the latter
being due to local conservation laws in integrable models
\cite{zotos1997, rosch2000, sirker11, prosen11, prosen13}. These
unusual properties have been speculated to be related to the  huge
magnetic thermal conductivities observed in 1D quantum magnets
\cite{sologubenko07, hess07, hlubek10} and may have potential
applications in signal propagation in artificial 1D systems on
surfaces \cite{menzel12} or for spintronics applications
\cite{trauzettel08, hogdalem11}. Generic non-integrable 1D systems
are believed to exhibit no ballistic dynamics \cite{hm03, hm04,
jung2006, zotos2004} and presumably diffusive transport  (see Refs.\
\onlinecite{karrasch13b, znidaric13, steinigeweg2013-2,
steinigeweg2007-1} for possible exceptions or corrections beyond
diffusion). Recent experimental studies on 1D quantum magnets have
set out to elucidate spin diffusion using $\mu$sr and NMR
\cite{maeter2012, xiao2014}.

More recently, it has become possible to address qualitative aspects
of mass transport in experiments with ultra-cold quantum gases in
optical lattices \cite{schneider12, ronzheimer13, fukuhara13},
based on the realization of Bose- and Fermi-Hubbard models in these
systems \cite{bloch08}. Sudden expansion experiments, using the
release of a trapped gas of atoms into an empty and homogeneous
optical lattice, suggest that mass transport in two-dimensional
Hubbard models is diffusive for both bosons and fermions
\cite{schneider12, ronzheimer13}. Bosons in 1D subject to infinitely
strong interactions, called hard-core bosons, however, are
integrable via the exact mapping to non-interacting fermions
\cite{cazalilla11} and the results of \cite{ronzheimer13} establish
an unambiguous experimental realization of ballistic dynamics in an
integrable 1D system, rendering this a suitable starting point for
future studies. Moreover, hard-core bosons are equivalent to
spin-$1/2$ XX models \cite{cazalilla11}, thus providing a connection
to research on the transport properties of quantum magnets.

An important question concerns the effect of integrability breaking
on transport properties. In quantum gas experiments, a
straightforward way to break integrability is to induce an
inter-chain coupling and indeed, experimental results for sudden
expansions in the 1D-2D crossover of interacting bosons indicate a
rapid emergence of diffusive-like behavior upon increasing the
inter-chain coupling \cite{ronzheimer13}.

As an alternative to the dimensional crossover, one can consider two
coupled chains, i.e., a ladder, which can easily be realized in
optical lattices using superlattices \cite{foelling07}. The ladder
is accessible to state-of-the-art numerical methods, while for
two-dimensional systems, there are no reliable approaches. First
studies of the dynamics of hard-core bosons on a ladder geometry in
the sudden expansion \cite{vidmar13} or for wave-packet dynamics
\cite{karrasch2014} indicate diffusive dynamics for sufficiently
large inter-chain coupling, yet a rigorous analysis of ballistic and
diffusive contributions based on linear response theory is  lacking.

In this work, we address precisely this question using the
spin-$1/2$ XX ladder. By exploiting the concept of dynamical quantum
typicality \cite{hams2000, bartsch2009, elsayed2013,
steinigeweg2014-1, steinigeweg2014-2}, we are able to study ladders
with up to $N \leq 36$ spins, going beyond the range of  exact
diagonalization. In addition, we can reach the long time scales
required to analyze ballistic contributions
\cite{steinigeweg2014-1}. In the high-temperature limit, we first
demonstrate the absence of ballistic contributions and we show that
the model realizes standard diffusive dynamics, i.e., there is a
single relaxation time. We further compute the diffusion constant as
a function of the inter-chain coupling and, as another main result,
we identify three regimes characterized by qualitatively different
time-dependencies of current autocorrelation functions. Finally, we
discuss a possible experiment with quantum gases that could put our
theoretical predictions to a test.

The plan of this paper is the following: In Sec.\ \ref{sec:model},
we introduce the Hamiltonian and define the quantities of interest,
namely, the spin-current autocorrelation function, the Drude weight,
and the diffusion constant. In Sec.\ \ref{sec:method}, we briefly
discuss the numerical method, which is based on the concept of dynamical
typicality. Section \ref{sec:results} contains our main results for
the time dependence of the spin-current autocorrelation, the finite-size
scaling of the ballistic contribution, and the diffusion constant. We
also discuss our results in the context of recent quantum-gas experiments
with interacting bosons in optical lattices and make a proposal for a
future experiment designed to observe our predictions. Our conclusions
are presented in Sec.\ \ref{sec:conclusion}.

\section{Model and definitions}
\label{sec:model}

 We study spin-current dynamics in an XX ladder of length
$N/2$ with periodic boundary conditions, where $N$ is the number of
sites. The Hamiltonian $H = J_\parallel H_\parallel + J_\perp
H_\perp$ consists of a leg part $H_\parallel$ and rung part
$H_\perp$, given by ($\hbar=1$)
\begin{eqnarray}
&& H_\parallel = \!\! \sum_{i=1}^{N/2} \sum_{k=1}^2 (S^x_{i,k}
S^x_{i+1,k} + S^y_{i,k} S^y_{i+1,k}) \, , \nonumber \\
&& H_\perp =  \sum_{i=1}^{N/2} (S^x_{i,1} S^x_{i,2} + S^y_{i,1}
S^y_{i,2}) \, , \label{model}
\end{eqnarray}
where $S_{i,k}^{x,y}$ are spin-1/2 operators at site $(i,k)$,
$J_\parallel > 0$ is the antiferromagnetic exchange coupling
constant along the legs, and $J_\perp=r \,J_\parallel > 0$ is the
strength of the rung coupling. While the XX ladder splits into two
integrable XX chains of free Jordan-Wigner fermions for $r = 0$, it
simplifies to a set of uncoupled dimers for $r \to \infty$. In the
case of $r \neq 0$, the XX ladder is non-integrable and the
Jordan-Wigner transformation maps hard-core bosons to interacting
fermions \cite{vidmar13}. In general, the model in Eq.\
(\ref{model}) preserves the total magnetization $S^z$ and is
invariant under translations with periodic boundary conditions. We
take into account the full Hilbert space with $d=2^N$ states and
focus on the case for which $\langle S^z \rangle = 0$, see Sec.\
\ref{sec:tempmag}.

The longitudinal spin current is defined via the continuity
equation and has the form
\begin{equation}
j = J_\parallel \sum_{i=1}^{N/2} \sum_{k=1}^2 (S_{i,k}^x S_{i+1,k}^y -
S_{i,k}^y S_{i+1,k}^x) \, . \label{current}
\end{equation}
$\lbrack H,j\rbrack=0$ holds only at $r=0$. Within linear response
theory, we are interested in the current autocorrelation function at
inverse temperatures $\beta = 1/T$ ($k_B = 1$),
\begin{equation}
C(t) = \text{Re} \frac{\langle j(t) \, j \rangle}{N} =
\text{Re} \frac{\text{Tr} \{e^{-\beta H} j(t) \, j \}}{N \,
\text{Tr}\{e^{-\beta H}\}} \, , \label{exact}
\end{equation}
where the time argument of $j$ has to be understood w.r.t.\ the
Heisenberg picture, $j = j(0)$, and $C(0) = J_\parallel^2/8$ in the
limit $\beta \to 0$. From this autocorrelation we obtain the two
central quantities
\begin{equation}
\overline{C} = \frac{1}{t_2-t_1} \int_{t_1}^{t_2} \! \text{d}t \,
C(t) \, , \quad D  = \frac{1}{\chi} \int_0^{t_2} \! \text{d}t \,
C(t) \label{quantities}
\end{equation}
with $t_2 > t_1$ and $t_2 \to \infty$. The first quantity
$\overline{C}$ is the spin Drude weight, which, being the
non-decaying part of $C(t)$, signals ballistic transport
\cite{zotos1997}. The second quantity $D$ is the spin-diffusion
constant, well-defined for a vanishing $\overline{C}$ [and a
sufficiently fast decay of $C(t)$]. The prefactor $\chi$ is the
static susceptibility (per spin) and $\chi = 1/4$ as $\beta \to 0$.
The numerical calculation of the two quantities in Eq.\
(\ref{quantities}) is feasible by choosing finite but sufficiently
long times $t_1$ and $t_2$, where $C(t)$ has already decayed to its
final value, and $t_2\gg t_1$.

\section{Numerical method: Dynamical typicality}
\label{sec:method}

Our numerical method relies on replacing the
trace $\text{Tr}\{\bullet\}=\sum_n \langle n | \bullet | n \rangle$
in Eq.\ (\ref{exact}) by a scalar product involving a single pure
state $| \psi \rangle$. More precisely, following the concept of
quantum typicality, we draw $| \psi \rangle$ at random according to
a probability distribution that is invariant under all possible
unitary transformations in Hilbert space (Haar measure). Using a
so-constructed $| \psi \rangle$ and abbreviating $| \psi_\beta
\rangle = e^{-\beta H/2} | \psi \rangle$, the autocorrelation
function in Eq.\ (\ref{exact}) is approximated by \cite{hams2000,
bartsch2009, elsayed2013, steinigeweg2014-1, steinigeweg2014-2}
\begin{equation}
C(t) \approx \text{Re}\frac{\langle \psi_\beta | j(t) \, j |
\psi_\beta \rangle}{N \, \langle \psi_\beta | \psi_\beta \rangle} \,
, \label{approximative}
\end{equation}
the approximation becoming more accurate as the dimension of the
Hilbert space increases~\cite{JIN10X, steinigeweg2014-1}.

The salient feature of Eq.~(\ref{approximative}) is that it can be
calculated numerically without diagonalization of the Hamiltonian.
To this end one has to introduce two pure states: The first reads $|
\Phi_\beta(t) \rangle = e^{-\imath H t -\beta H/2} \, | \psi
\rangle$ and the second is $| \varphi_\beta(t) \rangle = e^{-\imath
H t} \, j \, e^{-\beta H/2} \, |\psi \rangle$. Then,
\begin{equation}
\langle \psi_\beta | j(t) \, j | \psi_\beta \rangle = \langle
\Phi_\beta(t) | j | \varphi_\beta(t) \rangle \, .
\end{equation}
The dependence of the two states on $t$ and $\beta $ is calculated
numerically by a massively parallel implementation of a
Suzuki-Trotter product formula or Chebyshev polynomial algorithm.
This allows us to study quantum systems with as many as $N=36$ spins
[Hilbert-space dimension $d = {\cal O}(10^{11})$], although we do
not exploit symmetries of Eqs.\ (\ref{model}) and (\ref{current}) at
present \cite{steinigeweg2014-1}.

\begin{figure}[t]
\includegraphics[width=0.90\columnwidth]{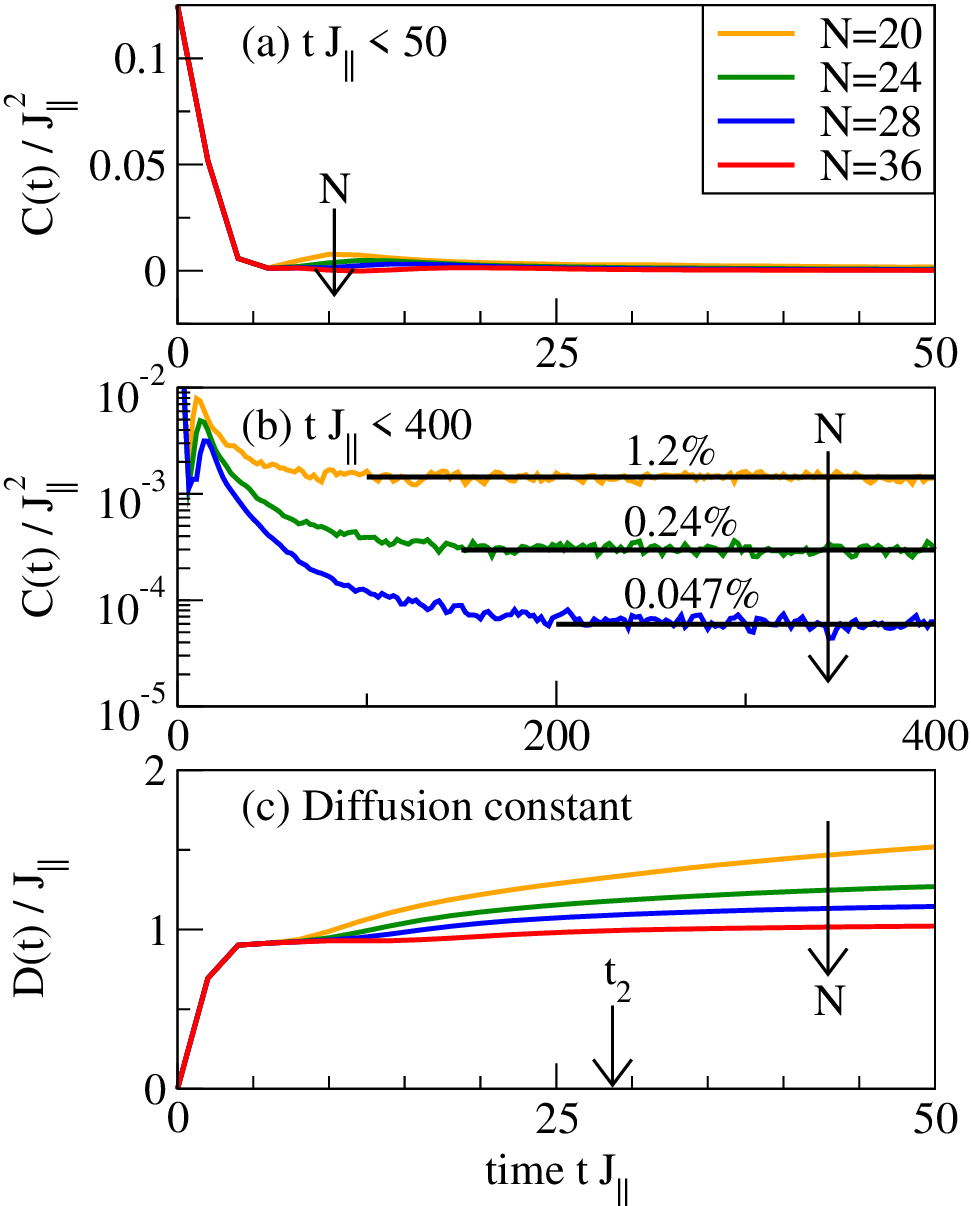}
\caption{(color online) Spin-current autocorrelation function $C(t)$
for $r=J_\perp/J_\parallel=1$ and $\beta=0$: (a) Relaxation curve
for large systems $N \leq 36$ and times $t \, J_\parallel \leq 50$;
(b) A saturation at a very small Drude weight is only visible in a
semi-log plot of (a) for very long times $t \, J_\parallel \gg 50$;
(c) Since the Drude weight $\overline C \lesssim \mathcal{O}(1\%)$,
$D(t) \to \text{const.}$ for large times.} \label{Fig1}
\end{figure}

\begin{figure}[b]
\includegraphics[width=0.90\columnwidth]{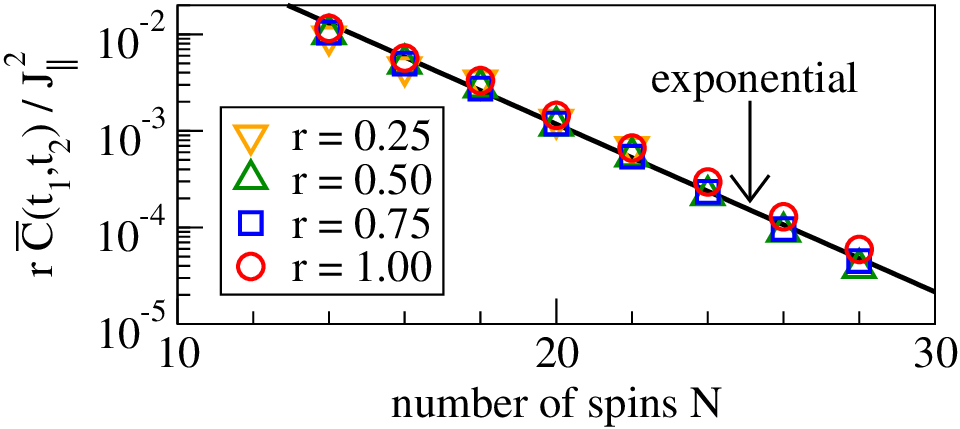}
\caption{(color online) High-temperature spin Drude weight
$\overline{C}(t_1,t_2)$, extracted at very long times $[t_1
J_\parallel, t_2 J_\parallel] = [300, 400]$, for different coupling
ratios $r=J_\perp/J_\parallel = 0.25$, $0.5$, $0.75$, and $1$
(symbols). The finite-size scaling follows an exponential $A(r) \,
e^{-\gamma N}$ (solid line, $\gamma \neq \gamma(r)$ and
$A(r) = 1/r$), suggesting $\overline C \to 0$ for $N \to \infty$.}
\label{Fig2}
\end{figure}

\begin{figure}[t]
\includegraphics[width=0.90\columnwidth]{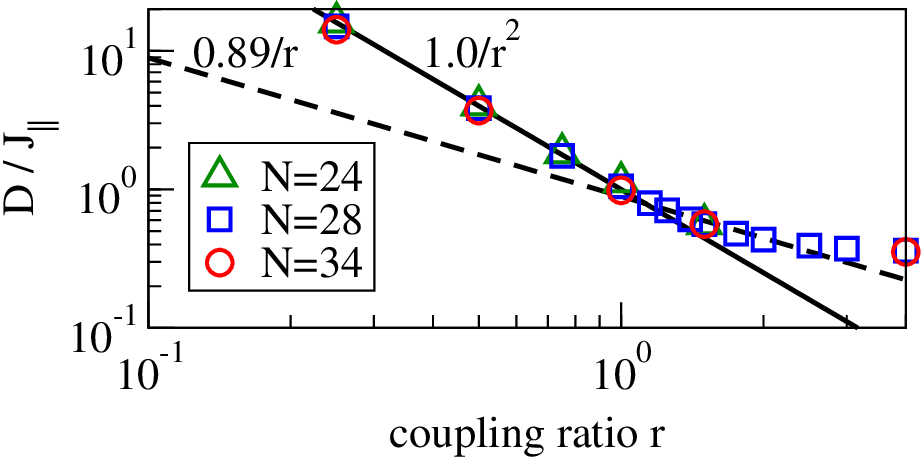}
\caption{(color online) Spin-diffusion constant $D$ versus $r =
J_\perp/J_\parallel$ for $\beta \to 0$. There are apparently three
scaling regimes: (i) $r \ll 1$: $D \propto 1/r^2$, (ii) $1 \lesssim
r \lesssim 2$: $D \propto 1/r$, and (iii) $r \gg 1$: $D =
\text{const}$. The $1/r$ curve results from  Eq.\ (\ref{Gaussian}),
see text.} \label{Fig3}
\end{figure}

\section{Results}
\label{sec:results}

\subsection{Time dependence of current autocorrelations}

We begin with high temperatures $\beta \to 0$ and an
intermediate rung interaction strength $r=1$. Figure \ref{Fig1}(a)
summarizes our numerical results for $C(t)$ for different system
sizes $N=20$, $24$, $28$, and $36$. Clearly, $C(t)$ rapidly decays
towards zero for all $N$, with almost no finite-size effects visible
in the lin-lin plot. Particularly, for all $N$ depicted, there is no
signature of a dissipationless contribution of $C(t)$ for times $t
\, J_\parallel \leq 50$. To illustrate the existence of such a
contribution, Fig.\ \ref{Fig1}(b) shows a semi-log plot of $C(t)$ up
to times $t \, J_\parallel \leq 400$. Although a dissipationless
contribution becomes visible, it amounts to only $1\%$ of the
initial value $C(0)$ for $N=20$ and systematically decreases further
when $N$ is increased, taking a tiny value $\ll 1\%$ for $N=28$.
Note that we do not determine $\bar{C}$ for larger $N$ since, for
such $N$, the computational effort is unreasonably high for the long
times $t \, J_\parallel > 400$ required.

\subsection{Absence of ballistic contributions for large $N$}

In Fig.\ \ref{Fig2} we provide a detailed finite-size analysis of
the non-decaying contribution, based on system sizes where this
contribution can be extracted from the  long-time window $[t_1
J_\parallel, t_2 J_\parallel] = [300, 400]$, see Fig.\ \ref{Fig1}(b)
as well as the definition of $\overline{C}$ in Eq.\
(\ref{quantities}). Using a log-lin plot unveils an exponential
decrease with system size, over more than two orders of magnitude.
Certainly, this kind of decrease may be expected for a highly
non-integrable model \cite{steinigeweg2014-2} but we observe this
scaling for various rung couplings $r= 0.25$, $0.5$, $0.75$, and
$1$. What is more, the exponent turns out to be practically
independent of $r$ while the amplitude scales roughly as $\propto
1/r$. Based on these results, we conclude that, for $r>0$, the Drude
weight vanishes in the thermodynamic limit. Compared to earlier
studies of transport in gapped 1D spin systems \cite{zotos2004,
jung2006, hm03, hm04}, we resolve a particularly clean exponential
and fast decay of the Drude weight.

\subsection{Diffusion constant}

Since the Drude weight vanishes, the central quantity of interest is
the diffusion constant. In fact, we are able to calculate the
diffusion constant even quantitatively using large systems, due to
the tiny non-decaying contribution for such systems. Still, we have
to choose a finite time $t_2$ for the evaluation of $D$ in Eq.\
(\ref{quantities}). In praxis, we determine the decay time $\tau$,
where $C(\tau)/C(0) = 1/e$, and calculate $D$ for $t_2 = 5.5 \tau \,
\gg \tau$. For instance, from the data shown for $r=1$ in Fig.\
\ref{Fig1}(c), we get $t_2 \, J_\parallel \approx 28$ and therefore,
a reasonable choice of $t_2$ with little finite-size effects for
large $N$. Note that we cannot choose extremely long $t_2$, which
would artificially blow up tiny non-decaying contributions or
include other finite-size effects. In Fig.\ \ref{Fig3} we depict
the resulting quantitative values of the diffusion constant as a
function of the rung coupling $r$. Values for different $N$
exhibit little finite-size effects for all $r$. The log-log
plot clearly unveils several regimes with a power-law dependence
of $D$ on $r$. More precisely, we observe three qualitatively
different regimes: (i) $r \ll 1$: $D \propto 1/r^2$, (ii) $1
\lesssim r \lesssim 2$: $D \propto 1/r$, and (iii) $r \gg 1$:
$D = \text{const}$. The intermediate regime (ii) is notably
much narrower than regimes (i) and (iii), yet distinct by the $D
\propto 1/r$ scaling. Note that in the XXZ chain similar regimes
appear as a function of the exchange anisotropy
\cite{steinigeweg2010, steinigeweg2011-1, karrasch2014}.

\begin{figure}[t]
\includegraphics[width=0.90\columnwidth]{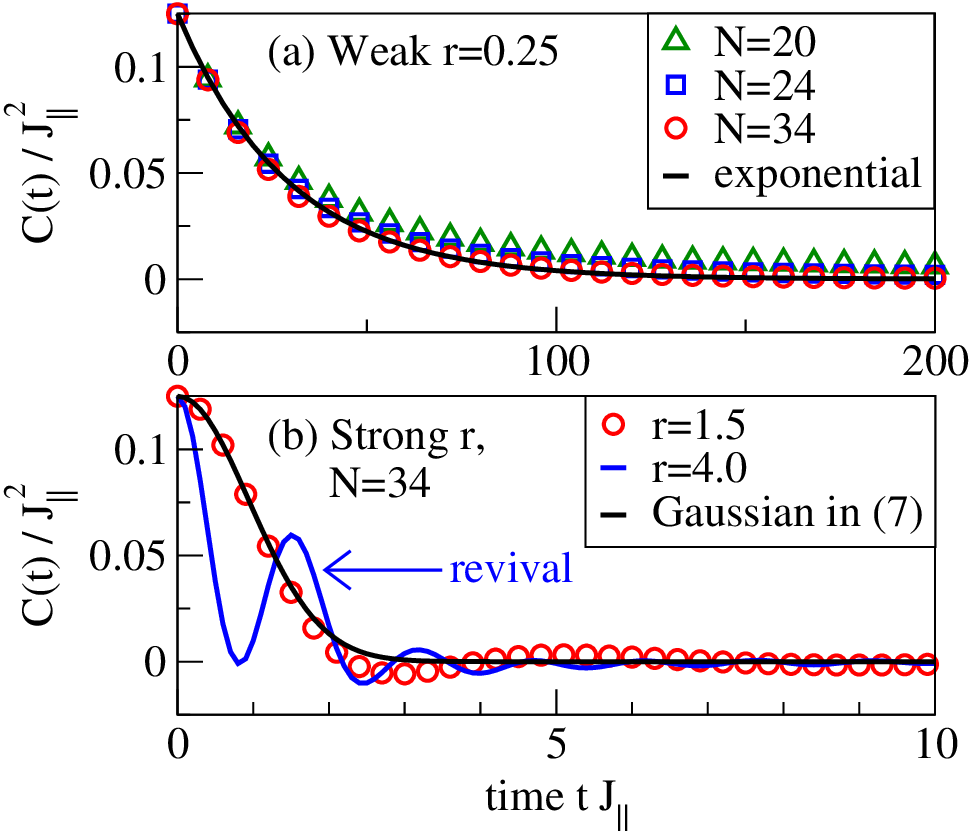}
\caption{(color online) Spin-current autocorrelation function $C(t)$
for  $\beta \to 0$: qualitatively different regimes, depending on
$r = J_\perp/J_\parallel$. (a) Weak-$r$ regime:  $C(t)$ decays
exponentially, resulting in $D \propto 1/r^2$ as expected from
perturbation theory. (b) Strong-$r$ regime: The decay curve agrees
with the Gaussian prediction of Eq.\ (\ref{Gaussian}), in line with
the generic behavior $D \propto 1/r$ suggested in Refs.\
\onlinecite{steinigeweg2010, steinigeweg2011-1}. For large $r$,
revivals occur resulting in $D = \text{const.}$} \label{Fig4}
\end{figure}

\begin{figure}[b]
\includegraphics[width=0.90\columnwidth]{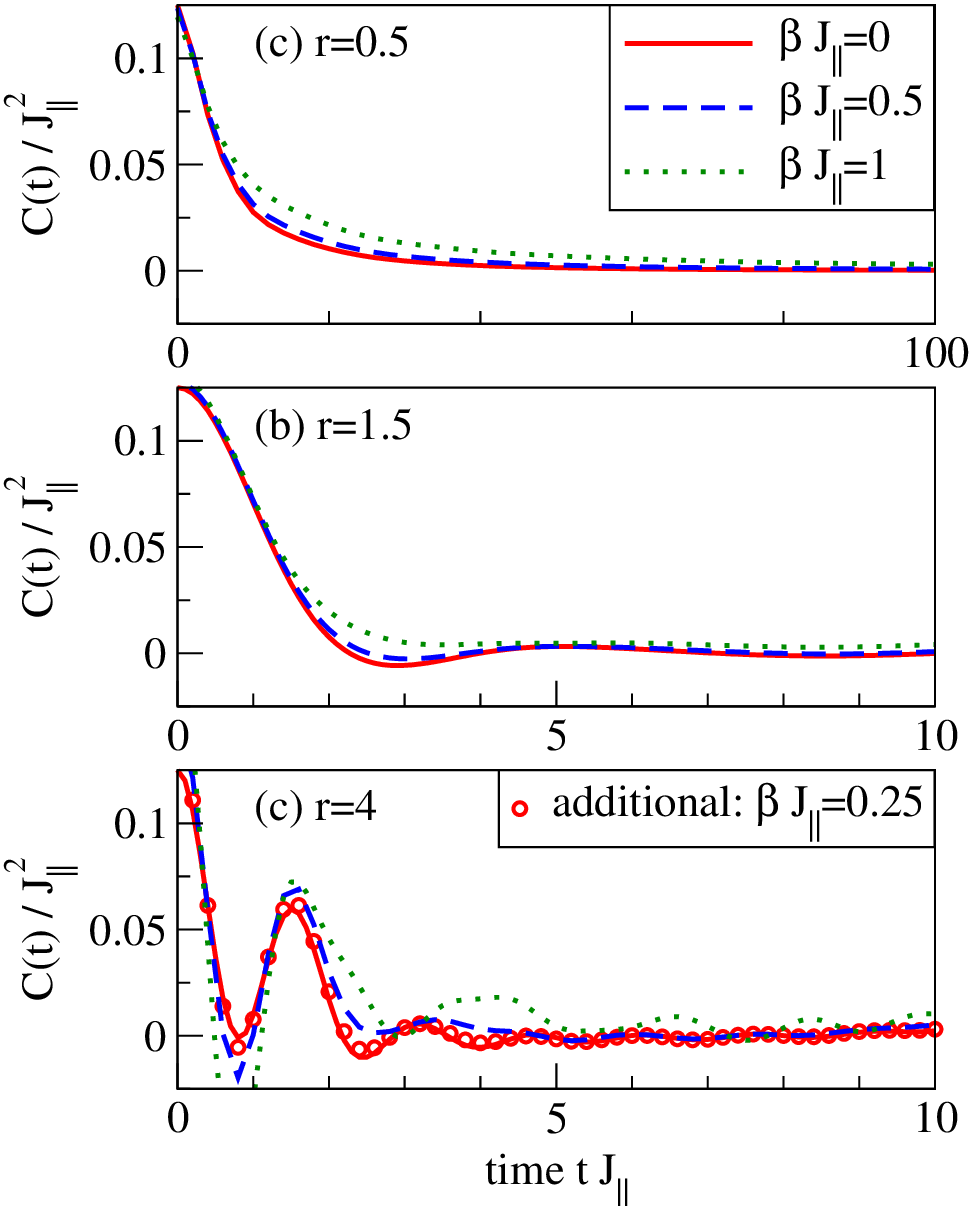}
\caption{(color online) Spin-current autocorrelation function $C(t)$
in the XX ladder for rung couplings (a) $r=0.5$,  (b) $r=1.5$, (c)
$r=4$ and different temperatures $\beta \, J_\parallel \leq 1$. In
all cases, $N=28$. Apparently, $C(t)$ is qualitatively the same for
temperatures down to $\beta \, J_\parallel \sim 0.5$.}
\label{Fig5}
\end{figure}

To gain insight into the origin of the scaling of $D$ with $r$, we
consider the time dependence of the spin-current autocorrelation
function $C(t)$ in more detail. In Fig.\ \ref{Fig4}(a) we show
$C(t)$ for a weak rung coupling $r = 0.25$. Evidently, the time
dependence of $C(t)$ is well described by a simple exponential
relaxation, implying standard diffusion. Due to this exponential
relaxation and the scaling $D \propto 1/r^2$ in Fig.\ \ref{Fig3},
the weak $r \ll 1$ regime turns out to be a conventional
perturbative regime \cite{steinigeweg2010, steinigeweg2011-1}. For
$r > 1$ the behavior changes qualitatively. In Fig.\ \ref{Fig4}(b)
we depict $C(t)$ for $r=1.5$. Here the exponential relaxation turns
into a Gaussian decay. This kind of decay, and particularly the
scaling $D \propto 1/r$ evident from  Fig.\ \ref{Fig3}, is in line
with the generic behavior suggested in \cite{steinigeweg2010,
steinigeweg2011-1} for the case of strong perturbations. In fact,
according to Ref.\ \onlinecite{steinigeweg2011-1}, one expects at
high temperatures $\beta\to 0$
\begin{equation}
C(t) = C(0) \, e^{-r^2 \gamma t^2} \, , \quad \gamma =
\frac{\text{Tr} \{\imath [j,H_\perp]^2 \}}{\text{Tr} \{ j^2 \}} =
\frac{1}{4} \label{Gaussian}
\end{equation}
and therefore, $D = C(0) \sqrt{\pi}/(2 r \chi \sqrt{\gamma}) \approx
0.89/r$. The prediction of Eq.\ (\ref{Gaussian}) is in good
agreement with the numerical data for $C(t)$ at $r = 1.5$ shown in
Fig.\ \ref{Fig4}(b). However, it does not account for  possible
revivals of $C(t)$ that occur in our case because of the band-like
spectrum that emerges in the limit of strong rung dimers for $r \to
\infty$. In this limit ($r$ large but finite), transport is mediated
by the triplet excitations above the dimer ground state
\cite{hess07}. In Fig.\ \ref{Fig4}(b) we illustrate the onset of
such revivals for $r=4$. These revivals define the third regime with
$D = \text{const.}$ shown in Fig.\ \ref{Fig3}, in analogy to the
spin-$1/2$ XXZ chain, where a similar behavior emerges in the
vicinity of  the Ising limit \cite{karrasch2014}. The observation of
diffusive transport with a single relaxation time and the
identification of the three scaling regimes characterized by
qualitatively different decays of current autocorrelations
constitute main results of this work.

\subsection{Finite temperatures and finite magnetization}
\label{sec:tempmag}

\begin{figure}[t]
\includegraphics[width=0.90\columnwidth]{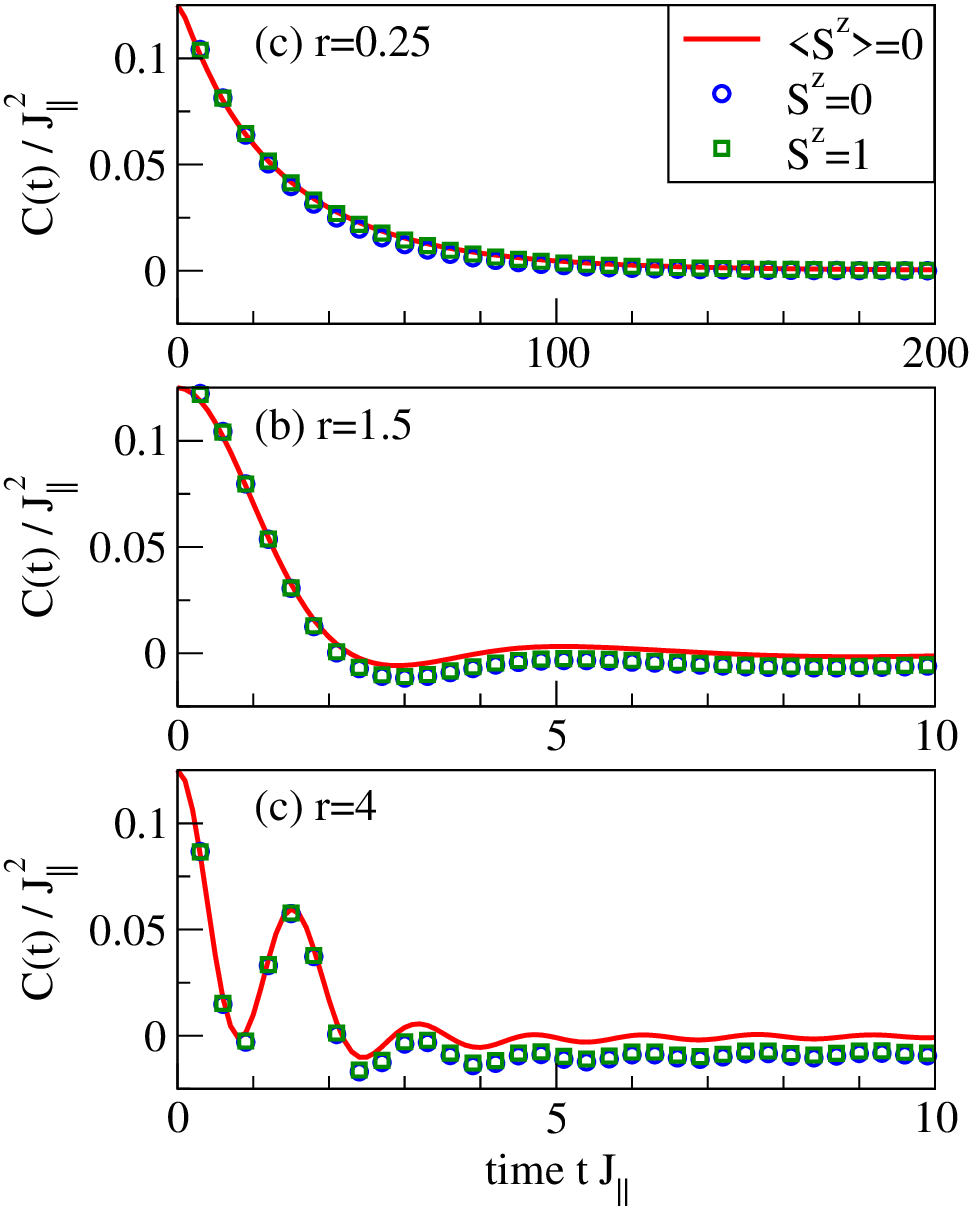}
\caption{(color online) Spin-current autocorrelation function $C(t)$
in the XX ladder for rung couplings (a) $r=0.25$,  (b) $r=1.5$, (c)
$r=4$ in the high-temperature limit $\beta \to 0$ and $\langle S^z
\rangle = 0$, $S^z = 0$, $S^z = 1$. In all cases, $N=34$. Apparently,
the cases of half filling $S^z = 0$ and almost half filling $S^z = 1$
are practically identical for finite $N$ already. No significant
difference to $\langle S^z \rangle = 0$ is visible for $r=0.25$,
while differences for larger $r$ are finite-size effects, as
illustrated in Fig.\ \ref{Fig7}.}
\label{Fig6}
\end{figure}

\begin{figure}[t]
\includegraphics[width=0.90\columnwidth]{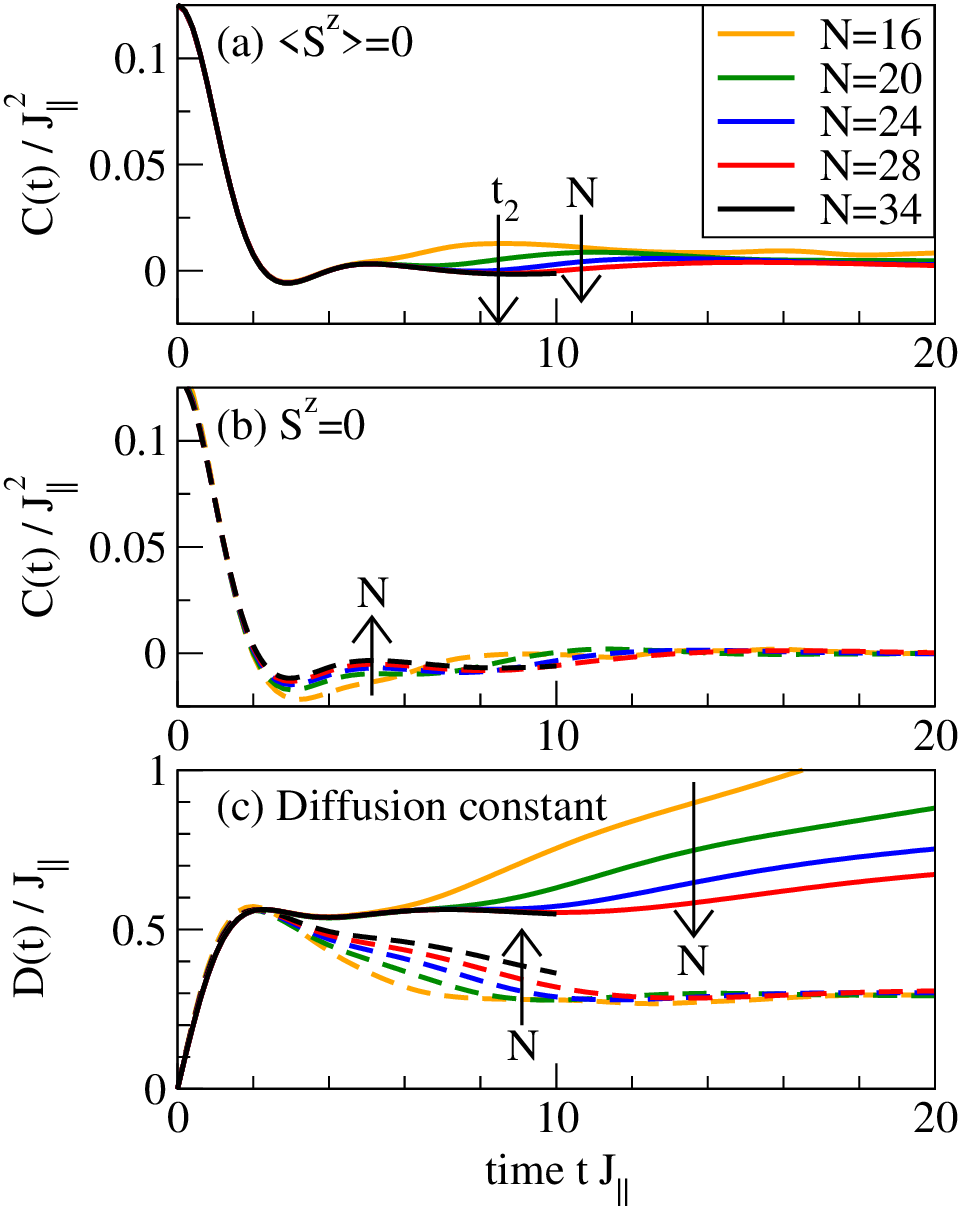}
\caption{(color online) Spin-current autocorrelation function $C(t)$
in the XX ladder for (a) $\langle S^z \rangle= 0$, (b) $S^z=0$ for a
strong rung coupling $r=J_\perp/J_\parallel=1.5$ in the high-temperature
limit $\beta \to 0$. (c) The resulting time-dependent diffusion coefficient,
as extracted from $C(t)$ depicted in (a), (b). Although finite-size results
for $\langle S^z \rangle=0$ and $S^z=0$ differ from each other, they seem
to converge to the same value in the thermodynamic limit. Apparently, the
convergence of the $\langle S^z \rangle=0$ data is much faster in time and there are
no finite-size effects up to times $t \, J_\parallel \sim 10$ comparing
$N = 28$ and $N=34$. Hence, our choice of $t_2 \, J_\parallel = 8.3$ for
extracting $D$ from $D(t)$ is reasonable.}
\label{Fig7}
\end{figure}

The qualitative dependence of $D$ on $r$ depends on temperature
(see, e.g., Ref.\ \onlinecite{damle05} for a theory of diffusion
in 1D gapped quantum magnets at low $T$). In Fig.\ \ref{Fig5}, we
check for the three different $r$ regimes that the qualitative
decay of $C(t)$ does not change down to $T / J_\parallel \sim 2$
and hence, it is reasonable to expect no qualitative changes in the
$r$ dependence of $D$. In Fig.\ \ref{Fig6}, we check that a finite
magnetization $S^z \sim 0$ does not change the picture either. We
note that the small differences between $\langle S^z \rangle = 0$,
$S^z = 0$, and $S^z = 1$ visible in Fig.\ \ref{Fig6} are finite-size
effects and vanish in the thermodynamic limit $N \to \infty$, see Fig.\
\ref{Fig7}. Remarkably, the convergence to that limit is the fastest
for $\langle S^z \rangle = 0$, which is the reason for focusing
on this ensemble in our paper.

\subsection{Connection to quantum gas experiments}

With respect to the recent sudden expansion experiment
\cite{ronzheimer13} of strongly interacting bosons on coupled
chains, we have here provided theoretical evidence that such
systems indeed exhibit diffusive dynamics. Finally, we describe an
experiment with cold quantum gases, in which our quantitative
results for the diffusion constant could be verified. Spin-1/2 XX
models can be realized with a single-component Bose gas in an
optical lattice in the limit of infinitely strong repulsive on-site
interactions \cite{paredes04, ronzheimer13}. In order to probe
diffusion, one would desire a homogeneous background density with
half a particle per site, which could be accomplished by using a box
trap \cite{gaunt13} instead of harmonic trapping potentials. The
basic idea to measure $D$ is to induce a local perturbation in the
density, by, e.g., superimposing a dimple trap using methods along
the lines of \cite{friedman02}, and then to monitor the time
evolution of the density profile as a function of position. From
such information, one can extract the diffusion constant from the
time dependence of the variance, as demonstrated for 1D spin systems
\cite{karrasch2014, langer09}. In order to observe our predictions,
it is necessary to put the gas at sufficiently high temperatures.
This can be done by subjecting the gas to heating.

\section{Conclusions}
\label{sec:conclusion}

 We studied spin transport in the spin-$1/2$ XX
ladder at finite temperature. Within linear response theory and
using the concept of dynamical typicality, this simple and
experimentally realizable non-integrable model exhibits standard
diffusive dynamics. We found qualitatively different dependencies of
the spin-diffusion constant on the rung-interaction strength,
resulting from a crossover from exponential to Gaussian dissipation
at intermediate coupling strengths. Our results suggest that
strongly interacting bosons on coupled chains, studied
experimentally in \cite{ronzheimer13}, exhibit diffusive dynamics.

\section*{Acknowledgements}

We thank U.\ Schneider for helpful discussions. F.\ H.-M.\
acknowledges support from the DFG through FOR 912 via grant
HE-5242/2-2. The authors gratefully acknowledge the computing time
granted by the JARA-HPC Vergabegremium and provided on the JARA-HPC
Partition part of the supercomputer JUQUEEN at Forschungszentrum
J\"ulich.


%

\end{document}